\documentclass[superscriptaddress,twocolumn,showpacs,preprintnumbers,amsmath,amssymb,aps,prb]{revtex4}
\usepackage{graphicx}
\usepackage{dcolumn}
\usepackage{bm}
\begin{document}

\title{A Research-Based Curriculum for Teaching the Photoelectric Effect}


\author{S. B. McKagan}
\affiliation{JILA, University of Colorado, Boulder, CO 80309, USA}

\author{W. Handley}
\affiliation{Department of Physics, University of Colorado, Boulder, CO 80309, USA}

\author{K. K. Perkins}
\affiliation{Department of Physics, University of Colorado, Boulder, CO 80309, USA}

\author{C. E. Wieman}
\affiliation{Department of Physics, University of British Columbia, Vancouver, BC V6T 1Z1, CANADA}
\affiliation{JILA, University of Colorado, Boulder, CO 80309, USA}
\affiliation{Department of Physics, University of Colorado, Boulder, CO 80309, USA}

\date{May 16, 2007}

\begin{abstract}
Physics faculty consider the photoelectric effect important, but many erroneously believe it is easy for students to understand. We have developed curriculum on this topic including an interactive computer simulation, interactive lectures with peer instruction, and conceptual and mathematical homework problems.  Our curriculum addresses established student difficulties and is designed to achieve two learning goals, for students to be able to (1) correctly predict the results of photoelectric effect experiments , and (2) describe how these results lead to the photon model of light.  We designed two exam questions to test these learning goals.  Our instruction leads to better student mastery of the first goal than either traditional instruction or previous reformed instruction, with approximately $85\%$ of students correctly predicting the results of changes to the experimental conditions.  On the question designed to test the second goal, most students are able to correctly state both the observations made in the photoelectric effect experiment and the inferences that can be made from these observations, but are less successful in drawing a clear logical connection between the observations and inferences.  This is likely a symptom of a more general lack of the reasoning skills to logically draw inferences from observations.
\end{abstract}

\maketitle

\section{Introduction}

Understanding the photoelectric effect is a crucial step in understanding the particle nature of light, one of the foundations of quantum mechanics.  The photoelectric effect is a powerful tool to help students build an understanding of the photon model of light, and to probe their understanding of the photon model.  This topic, which may seem straightforward to physics professors, is treated only briefly in many courses in modern physics and quantum mechanics.  However, research shows that students have serious difficulties understanding even the most basic aspects of the photoelectric effect, such as the experimental set-up, experimental results, and implications about the nature of light~\cite{Steinberg1996a,Steinberg2000a,DeLeone2004a,Knight2004b}.  Thus, there is a need for better curriculum to address these difficulties.

To ensure that our curriculum is aligned with faculty goals and expectations when teaching the photoelectric effect, we conducted an online survey of faculty who had recently taught modern physics.  We received responses from 15 faculty members at 9 universities.

Everyone who responded to our survey said that it is important to teach the photoelectric effect.  Some typical explanations they gave for its importance are:
\begin{quotation}
The quantization of radiation was an important development in the advancement quantum theory. The beautiful and simple explanation of a puzzling effect is a rather powerful example of the efficacy of quantum principles and good old conservation laws.
\end{quotation}
\begin{quotation}
This is one of the most basic and foundational experiments, both historically and conceptually, for the nature of the photon and the concept of duality.
\end{quotation}

While all agreed on the importance of the photoelectric effect, the average faculty member spent about an hour of lecture time on this topic, and gave a few homework problems.  Thus, while most faculty members feel that the photoelectric effect is extremely important, they do not spend much time on it.  One third of our respondents described the photoelectric effect as ``simple.''  There seems to be a widespread perception among faculty that this topic is straightforward and can be understood by students with relatively little effort.

We also asked faculty to list their learning goals for their students when teaching this topic.  The majority of the responses were consistent with our two main goals, for students to be able to:
\begin{enumerate}
  \setlength{\itemsep}{1pt}
  \setlength{\parskip}{0pt}
  \setlength{\parsep}{0pt}
  \item correctly predict the results of experiments of the photoelectric effect, and
  \item describe how these results lead to the photon model of light.
\end{enumerate}
Most faculty (80\%) thought that they had achieved these goals.

\section{Previous Research on the Photoelectric Effect}

The results of physics education research (PER) tell a different story.  Faculty are overestimating both the simplicity of the photoelectric effect and their students' mastery.

Steinberg et al.~\cite{Steinberg1996a,Steinberg2000a} carried out studies of student learning of the photoelectric effect, consisting of interviews and analysis of exam questions.  They found that after standard instruction, many students did not have even a basic understanding of the experimental set-up or the implications of Einstein's explanation of it.  They summarized the specific difficulties they found as follows~\cite{Steinberg2000a}:
\begin{enumerate}
  \setlength{\itemsep}{1pt}
  \setlength{\parskip}{0pt}
  \setlength{\parsep}{0pt}
    \item a belief that V = IR applies to the photoelectric experiment
    \item an inability to differentiate between \textit{intensity of light (and hence photon flux)} and \textit{frequency of light (and hence photon energy)}
    \item a belief that a photon is a charged object
    \item an inability to make any prediction of an I-V graph for the photoelectric experiment
    \item an inability to give any explanation relating photons to the photoelectric effect
\end{enumerate}

In our own observations of students and analysis of homework and exam responses, we have observed evidence of all of these student difficulties except the belief that a photon is a charged object.

In response to these difficulties, Oberem and Steinberg developed a computer tutorial called Photoelectric Tutor (PT)~\cite{Oberem1999a}, that students completed on their own in about an hour after traditional instruction.  PT was designed mainly to address the first learning goal discussed in the introduction: for students to be able to correctly predict the results of experiments of the photoelectric effect.  As will be discussion in Section \ref{Assessment} (see Table \ref{exam1results}), this achieved substantial improvement, but not complete success, at achieving the desired goals.

De Leone and Oberem~\cite{DeLeone2004a} conducted further studies in a course using PT, confirming many of the findings of Steinberg et al. and demonstrating that many students lack a basic understanding of the classical model of light with which the results of the photoelectric effect are contrasted.  We have observed the same problem with our students, and have found that it is necessary to spend a significant amount of time reviewing the classical model of light.

In his book \textit{Five Easy Lessons}, Knight~\cite{Knight2004b} describes informal studies he has done on the photoelectric effect, which show that many students do not achieve the second learning goal of describing how the results of the photoelectric effect experiment lead the the photon model of light: ``When asked on an exam to explain how the photoelectric effect was inconsistent with classical physics, the majority of students wrote that the mere existence of the photoelectric effect violated classical physics.  Only a very small minority could articulate how the photon model succeeds where the classical model fails.''

In summary, research has shown that students lack much of the prerequisite knowledge of circuits and the classical model of light needed to understand the photoelectric effect, and that in traditional instruction, most students do not achieve either of the learning goals listed in the introduction.  While PT has helped students make progress on the first goal, predicting the results of experiments of the photoelectric effect, $60\%$ of these students were still unable to correctly predict the effect of changing the voltage.  We know of no previous research that has addressed whether reformed curriculum can help with the second goal, describing how these results lead to the photon model of light.

\section{Method of Instruction}
In Fall 2005 and Spring 2006, we reformed and taught a large lecture modern physics course for engineering majors.~\cite{McKagan2007a} The course had a relatively strong emphasis on reasoning development, model building, and real world applications.  In addition we implemented a variety of PER-based learning techniques, including concept tests with peer instruction, collaborative homework sessions, and interactive simulations.  In Fall 2006 and Spring 2007, this course was taught by another professor, also a member of the physics education research group, who used our curriculum.

Among many goals for the course was to address the difficulties described in previous research on the photoelectric effect and to go beyond the previous work on this subject towards achieving the learning goals discussed in the introduction.  Furthermore, we wanted to create a simulation and supporting materials that are freely available online, and thereby accessible to a wide audience.

\begin{figure}[b]
  \includegraphics[width=\columnwidth]{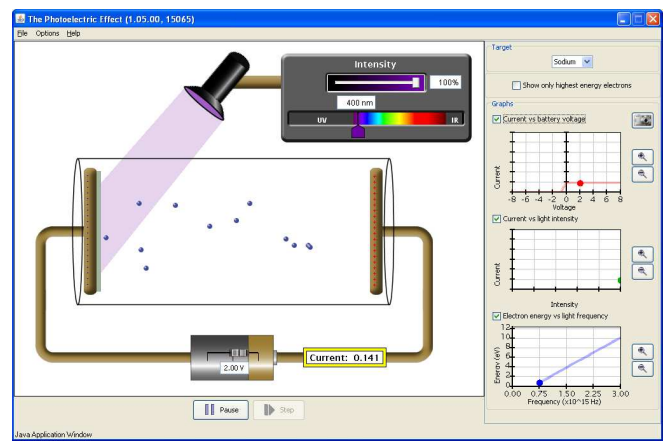}
  \caption{\textit{The Photoelectric Effect} simulation.}
  \label{simulation}
\end{figure}

Our resulting curriculum on the photoelectric effect, which is suitable for a sophomore level modern physics course, includes three 50 minute interactive lectures, conceptual and mathematical homework problems, and an interactive computer simulation.  \textit{The Photoelectric Effect} simulation, shown in Fig. \ref{simulation}, was designed as part of the Physics Education Technology Project (PhET), and is available for free download, along with many other simulations in introductory physics and quantum mechanics, from the PhET website~\cite{PhET}.  Our curriculum is available from both the PhET activities database and from our modern physics course archive~\cite{2130}.

The homework assignment includes multiple choice, calculation, and essay questions, all of which are centered around the simulation.  The homework asks students to predict the results of experiments that they can perform with the simulation, and to explain the reasons for, and significance of, these results.  According to students' self reports in Fall 2006, the average amount of time they spent on the photoelectric effect homework was 2 hours and 16 minutes, and about half of them worked with other students on it.

The first two lectures focus on understanding the basic experimental setup, results, and implications.  These lectures include content that would be included in any typical modern physics course, but have more emphasis on the necessary background knowledge of circuits and the classical wave model of light.  The third lecture is devoted to applications such as photomultiplier tubes, as well as details of how the electrons are bound in materials -- content that might not be included in a typical course.  These applications were included as part of our overall effort to make the course more relevant to the students by including real world applications.

\begin{figure}[htbp]
  \includegraphics[width=\columnwidth]{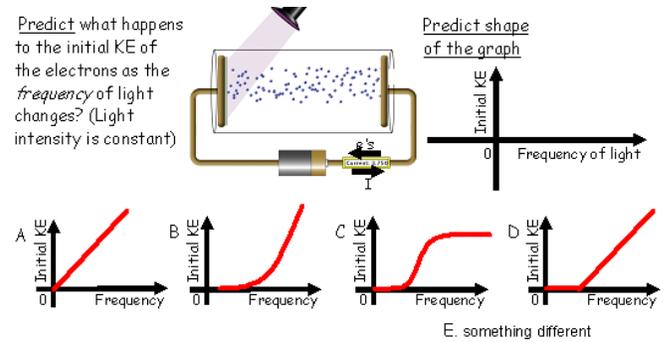}
  \caption{A sample clicker question used in class.}
  \label{clicker}
\end{figure}

\begin{figure}[htbp]
  \includegraphics[width=\columnwidth]{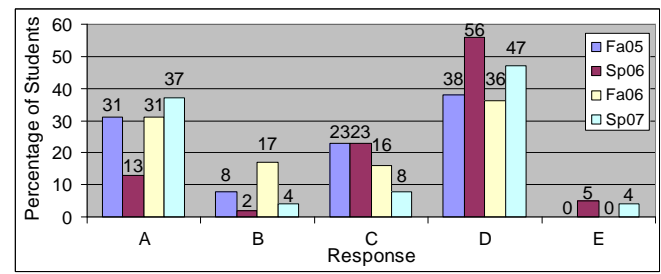}
  \caption{Percentage of students giving each response for clicker question shown in Fig. \ref{clicker}.}
  \label{clickerresults}
\end{figure}

Fig. \ref{clicker} shows a typical question from an interactive lecture demonstration, in which we ask students to predict the effect of changing the frequency of light on the kinetic energy of the electrons.  We collected student responses to such questions using clickers.  This question addresses several of the critical features of the photoelectric effect, including the linear dependence of kinetic energy on frequency and the existence of a cutoff frequency.  This is a difficult question, and only a third to a half of the students are able to answer that graph D is correct. (See Fig. \ref{clickerresults}) We find that group discussions are extremely productive for student learning in such questions, as the percentage able to answer correctly before discussion is much smaller.  For example, in Spring 2007, in which we collected student drawings before showing them the multiple choice answers, only $16\%$ drew something resembling graph D.

In the two spring semesters we asked students to draw the graph on a piece of paper before showing them the multiple choice options, and in the two fall semesters we showed the options right away.  We note that a significantly higher percentage of students ultimately answered the question correctly when we asked them to draw the graph first.  We hypothesize that this is because drawing a graph results in greater student engagement, since it is more active than choosing a graph from a list of options.

\section{Simulation}
We designed an interactive simulation to address the widespread student difficulties of the photoelectric effect discussed in the literature and apparent in our own studies.  The simulation allows students to control inputs such as light intensity, wavelength, and voltage, and to receive immediate feedback on the results of changes to the experimental set-up.  With proper guidance (in the form of interactive lecture demos and homework questions), students can use the simulation to construct a mental model of the experiment.  The simulation also allows students to interactively construct the graphs commonly found in textbooks, such as current vs. voltage, current vs. intensity, and electron energy vs. frequency.  By seeing these graphs created in real time as they change the controls on the experiment, students are able to see the relationship between the graphs and the experiment more clearly than when viewing static images.

\noindent\textbf{The Circuit:}
Previous research has shown that students have trouble understanding the circuit diagrams generally used in textbooks to represent the photoelectric effect.  Therefore we replace the circuit diagram with a cartoon-like picture of an actual experiment, as shown in Fig. \ref{simulation}.  We replace the variable voltage supply with a simple battery with a slider.  Our design is based on suggestions from previous research on the photoelectric effect~\cite{Knight2004b}, research showing that students learn best when you reduce their cognitive load by eliminating unnecessary details~\cite{Mayer2003a}, and our own research on simulation interface design and learning~\cite{Adams2007a,Adams2007b}.

\noindent\textbf{Electrons:}
One of the more controversial aspects of our simulation~\cite{Heron2004a} is that we show the electrons passing from one plate to the other, a feature that would not be visible in a real experiment.  We have observed that this aspect of the simulation is extremely useful in helping students visualize the effect of changing the voltage.  Students can see in a very concrete way that increasing the voltage accelerates the electrons and making the voltage negative decelerates them.

We have found that many students have difficulty understanding the relationship between current and electron speed.  In class and in problem-solving sessions, students often have heated debates about whether increasing the speed of the electrons leads to an increase in current.  The simulation is a critical tool in resolving these debates, because students can see upon close inspection that increasing the speed of the electrons does not increase the number arriving per second on the plate, and therefore does not increase the current.

The electrons also provide a compelling way to visualize the meaning of the stopping potential, a concept that previous research has shown to be extremely difficult for students~\cite{Steinberg1996a}.  In the simulation, when the battery voltage is tuned to exactly the stopping potential, students can see the electrons just make it to the opposite plate and then turn around.  The image of these electrons not quite making it seems to be a powerful one.  It often elicits laughter from students the first time they see it, and we observe them spontaneously describing it to instructors and other students long afterwards.

Some teachers who have seen the simulation have expressed concern that showing the electrons coming off the plate gives away the model of electron flow and therefore makes it ``too easy'' for students.  In response to this concern, we point out that instruction on the photoelectric effect normally assumes that students are already comfortable with the model of current as electron flow.  Research shows that this is not necessarily the case.  If students do not have a clear model of current, it will greatly impede their ability to understand the results of the photoelectric effect.  Our observations that students do poorly on many of the in-class concept questions and struggle with the homework on the photoelectric effect indicate that the simulation does not make learning of this topic ``too easy.''  Further, seeing a phenomenon on a computer screen is quite different from having an internal model of the phenomenon.  Our research shows that passively viewing an animation is not sufficient for building a mental model, and that students must interact with a simulation to learn from it.~\cite{Adams2007a}  Even with the animation of electrons, students still must expend considerable mental effort to formulate a useful mental model.

\noindent\textbf{Photons:}
In contrast to the electrons, we do not show individual photons, but instead represent light as a beam, an image that is consistent with either the wave or the particle model.  Because understanding the experimental basis of the photon model of light is the goal of instruction on the photoelectric effect, we want the simulation to aid students in constructing this model, rather than explicitly providing it.  The options menu in the simulation does allow instructors to show individual photons in place of the beam view.  (Our research shows that students rarely look in the options menu.~\cite{Adams2007b})  We find this view useful to illustrate the photon model \textit{after} we have already discussed how the evidence supports this model.

\noindent\textbf{Simplifications:}
Because the photoelectric effect experiment, like any real experiment, contains many subtle complications that are not relevant to the instructional goals, instructors must make decisions about which details to omit.  Most textbooks, for example, discuss the fact that electrons leave the plate with a range of energies, and thus the equation $KE \leq hf - \phi$ contains an inequality rather than an equal sign. On the other hand, most textbooks \emph{do not} discuss the fact that electrons leave the plate at different angles, although they do present current vs. voltage graphs that represent this behavior.  We have found that when instruction asks students to construct graphs based on a physical model, it is impossible to ignore either the range of energies or angles of the electrons, because if we don't bring them up, the students do.  This is not necessarily the case in traditional instruction.  More advanced issues that can safely be ignored in an introductory treatment are contact potential, thermionic emission, and reverse current~\cite{Preston1991a}.

\begin{figure}[htbp]
  \includegraphics[width=\columnwidth]{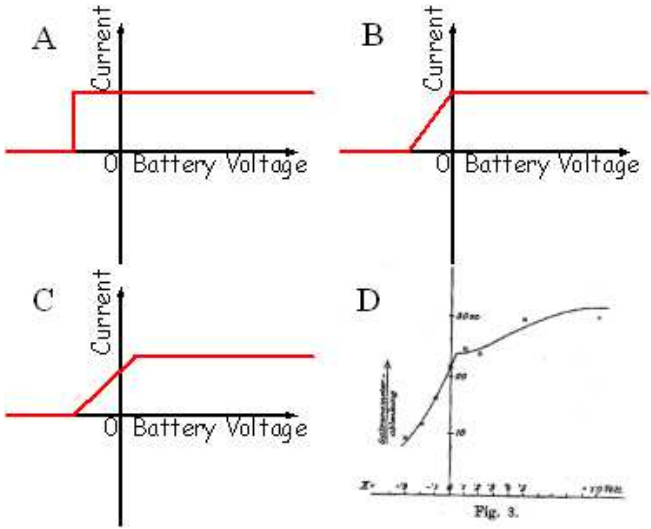}
  \caption{Current vs. voltage curves for (A) assuming all electrons leave perpendicular to the plate with the maximum kinetic energy, (B) assuming all electrons leave perpendicular to the plate with a range of energies, (C) assuming electrons leave the plate at a range of energies and with a range of angles, and (D) Lenard's experimental data~\cite{Lenard1902a}.}
  \label{IVcurves}
\end{figure}

The current vs. voltage graph, the central pedagogical tool in PT, provides a useful illustration of the key features of the photoelectric effect.  One must be very careful in drawing this graph, as the shape depends on the assumptions one makes, as can be seen in Fig. \ref{IVcurves}.  Most modern physics textbooks include a drawing that is some variation of Fig. \ref{IVcurves}C, although the straight diagonal line is generally replaced with some kind of curved line, the shape of which is different in every textbook.  (The details of the shape of this part of the curve depend on experimental issues that are beyond that scope of an introductory treatment of the photoelectric effect, and were impossible to determine from the original experimental data shown in Fig. \ref{IVcurves}D.)  In order to draw this curve, rather than the curves in Fig. \ref{IVcurves}A or \ref{IVcurves}B, one must understand that electrons leave the plate at both a range of energies and a range of angles.  In PT, if students draw a curve like Fig. \ref{IVcurves}A or \ref{IVcurves}B, the program carefully walks them through the reasoning they need to see why Fig. \ref{IVcurves}C is the correct curve.  However, a review of six of the most commonly used modern physics textbooks~\cite{Tipler2002a,Taylor2004a,Krane1996a,Harris1999a,Serway1989a,Beiser2003a} reveals that only two~\cite{Tipler2002a,Harris1999a} include even a brief parenthetical statement discussing \emph{why} electrons leave with a range of energies, and none mention the fact that electrons leave at a range of angles.

In the initial version of the simulation, we tried to reduce students' cognitive load by starting with a ``simple'' model in which all electrons were ejected with the same energy.  As students became more comfortable with this simple model, we then introduced the ``realistic'' model, in which electrons were ejected with a range of energies.  The simulation allowed students to switch between models with radio buttons labeled ``simple'' and ``realistic.''  This method led to a lot of confusion, as students recognized that there were many simplifying assumptions that could be made, and had trouble remembering which assumptions were included in the ``simple'' model.  They expended a lot of mental effort keeping track of which model we wanted them to use, effort that could have been focused on understanding the physics.

After the first semester of the course (Fall 2005), we replaced the ``simple'' and ``realistic'' radio buttons with a checkbox labeled ``show only highest energy electrons.''  It is unchecked by default, so that the simulation starts in the ``realistic'' model where electrons are ejected with a range of energies.  If this model is too overwhelming, the student or instructor can check ``show only highest energy electrons'' to simplify it, and it is clear exactly what simplification they are making.  We used this version in Spring 2006, and found that students seemed much less confused and the number of student complaints decreased.

We chose to make the simplifying assumption that all the electrons leave perpendicular to the plate, mainly because the benefit of varying the angles did not seem to be worth the cost.  Students often ask whether the electrons actually come off at different angles, and are generally willing to accept that this is just a simplification of the simulation.

A further issue in designing the simulation was how to define intensity as a function of frequency.  This issue does not typically arise in a real experiment because it is not physically practical to continuously tune the frequency of light while keeping the intensity constant, as one can do in the simulation.  The intensity of light is proportional to the energy of the beam, which is equal to the number of photons times the energy of each photon, $hf$.  Therefore, if you hold the intensity fixed and increase the frequency, the number of photons should decrease, since each photon contains more energy.  This is the way the simulation behaves, and it is physically correct, but it has caused some confusion among both students and instructors, who expect the number of photons to remain fixed (e.g. see Fig. 38.3 in Knight's textbook~\cite{Knight2004a}, which makes this error).  For instructors who prefer to make the simpler assumption that the number of photons remains fixed as the frequency changes, the options menu contains a choice to control the number of photons instead of the intensity.

\section{Unexpected Consequences}
One of the unexpected consequences of interactive engagement techniques, at least as implemented in our course, is that they encourage many difficult student questions about the implications of the material discussed in class.  We have found that when students are truly engaged in the material and are trying to make sense of it, they start asking many questions that are beyond the anticipated scope of the class, and sometimes even beyond the scope of knowledge of the instructors, including one Nobel laureate.  We point this out because it is important for instructors to be prepared for such tough questions.  This preparation includes having answers to questions one can anticipate, as well as the ability to think on one's feet and/or admit ignorance in response to the questions one cannot anticipate.  Here are some examples of questions students have asked during lectures on the photoelectric effect:
\begin{itemize}
  \setlength{\itemsep}{1pt}
  \setlength{\parskip}{0pt}
  \setlength{\parsep}{0pt}
    \item Wouldn't there be less current at low voltages because the electrons would fly off in different directions and not hit the other plate?
    \item How does the work function relate to where the element is in the periodic table?
    \item Why is intensity independent of frequency for light but not sound?
    \item Can two photons give energy to a single electron?
    \item How does the photon decide where it's going to hit the metal?
    \item Shouldn't those accelerating electrons be emitting light?
    \item Wouldn't kinetic energy of the electrons eventually level off because they can't go faster than the speed of light?
    \item Why does the light rip off electrons but not protons?
\end{itemize}

\section{Assessment}\label{Assessment}
To assess the effectiveness of our curriculum, we gave two exam questions that were designed to assess student learning of each of our two learning goals for the photoelectric effect.

\begin{figure}[t]
  \includegraphics[width=\columnwidth]{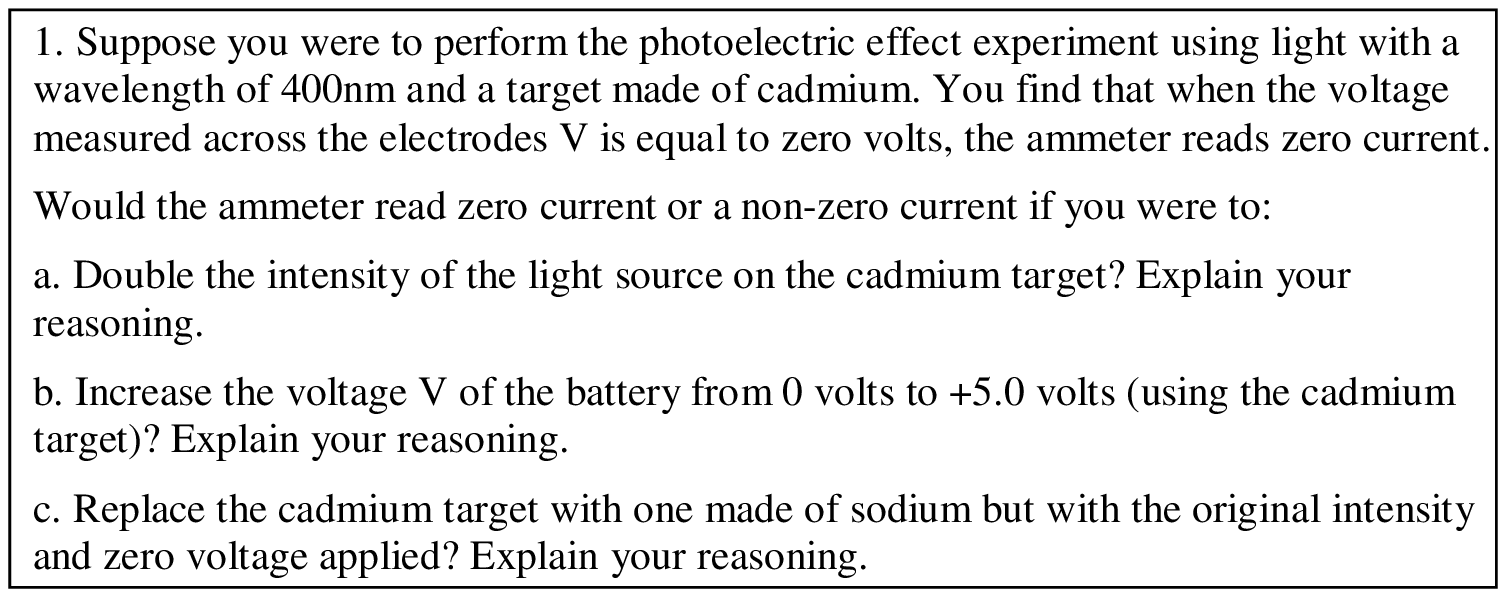}
  \caption{First exam question}
  \label{exam1}
\end{figure}

\noindent\textbf{First exam question:}
The first exam question, shown in Fig. \ref{exam1}, was adapted from the first exam question developed by Steinberg et al.~\cite{Steinberg1996a} to assess the effectiveness of PT.  We used this question so that we could compare our results to the results of this previous study.  We changed the order of the parts to make the question flow more smoothly, and changed the vocabulary slightly to be more consistent with the vocabulary used in our course (e.g. replacing the word ``cathode'' with ``target'' and ``potential difference'' with ``voltage'').  We changed the target material and numbers each semester, but otherwise the wording of the question was identical in each course.  We asked this question in three consecutive semesters.  This question provides an assessment of the first learning goal, for students to be able to correctly predict the results of experiments of the photoelectric effect.

\begin{table}[tbp]
\begin{tabular}{|r|r|r|r|r|}
\hline
           &     a &     b &     c &          N \\
\hline
UW non-PT Students &         40 &         20 &         65 &         26 \\
\hline
UW PT Students &         85 &         40 &         75 &         36 \\
\hline
     CU Fa05 &         87 &         85 &         91 &        189 \\
\hline
     CU Sp06 &         88 &         84 &         86 &        182 \\
\hline
     CU Fa06 &         78 &         77 &         90 &         94 \\
\hline
\end{tabular}
\caption{Percentage of students who answered each part of the first exam question correctly (including correct reasoning).  The first two lines are taken from Steinberg et al.~\cite{Steinberg1996a}, which has a different question order (Q2 for part a, Q3 for part b, and Q1 for part c).  N is the number of students included in the sample.}
\label{exam1results}
\end{table}

Table \ref{exam1results} shows the percentage of students who answered each part of the first exam question correctly in each semester of our course, as well as in the Steinberg et al. study.  While the students in the Steinberg et al. study who used PT were able to do significantly better than the students who did not use PT, they still did poorly on part b, the question about voltage.  Our students' scores were comparable to the PT students on part a, somewhat better on part c, and significantly better on part b.  This is in spite of the fact that our class size was much larger than the classes in the Steinberg et al. study.

We hypothesize that the reason our students did so much better on the voltage question is that the \textit{Photoelectric Effect} simulation provides such a compelling visual model for the effects of changing voltage, allowing the students to visualize the actual behavior of the electrons as the voltage is changed.

The most common error, made by nearly half ($42\%$) of the students who answered at least one part of the first exam question incorrectly, was the misapplication of Ohm's law, for example claiming that a voltage is necessary for current flow or to overcome the work function of the metal.  This was also the first student difficulty noted by Steinberg et al. (see Section II).  This further illustrates that understanding the effect of voltage is one of the most difficult aspects of the photoelectric effect for students.  The second most common error, made by $5\%$ of the students who answered at least one part incorrectly, was to claim that for the case where the photon energy is less than the work function, it is possible to eject electrons by increasing the intensity alone.  The remainder of the incorrect responses were not obviously categorizable.

It is also worth noting that there is no significant difference between the scores in the first two semesters, when the course was taught by the curriculum designers, and the third semester, when it was taught by another professor and the class size was significantly smaller.  This suggests that the success of the curriculum is not strongly dependent on the instructor or class size.

\begin{figure}[b]
  \includegraphics[width=\columnwidth]{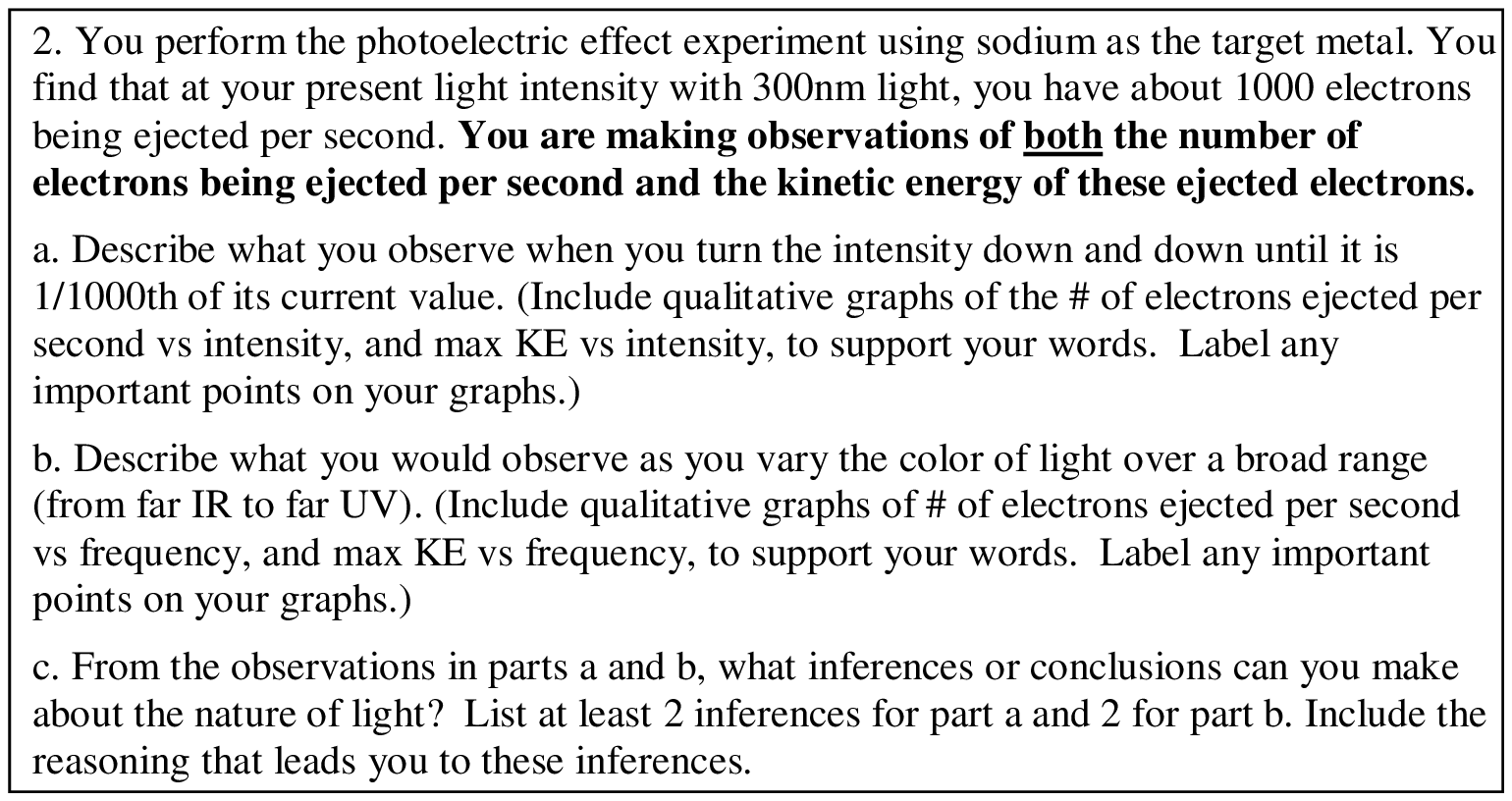}
  \caption{Second exam question (wording from Spring 2007).}
  \label{exam2}
\end{figure}

\noindent\textbf{Second exam question:}
The second exam question, shown in Fig. \ref{exam2}, was designed to measure whether students have achieved the second learning goal, to be able to describe how the experimental results lead to the photon model of light.  This question was asked all four semesters, although the wording was changed slightly each semester.  The request to draw specific graphs was added in Spring 2006, part c was separated from parts a and b in Fall 2006, and the request to ``list least 2 inferences for part a and 2 for part b'' was added in Spring 2007.

We note that part b of this question does not represent the historical order of events, since the observation of the effects of varying frequency were not made until \emph{after} Einstein predicted them, rather than being the basis of his inference of the particle model of light.~\cite{Arons1997a}  However, because making inferences from observations is quite difficult for students, we chose to give them more clues than Einstein had.

Because the analysis of the second question was more time-consuming than the first, we analyzed only a random subset of 47 students from each semester.  We analyzed student responses to the second exam question by recording the number and categories of correct observations and inferences that each student made, as well as by examining the reasoning that led from a specific observation to a valid inference.

\begin{figure}[bh]
  \includegraphics[width=\columnwidth]{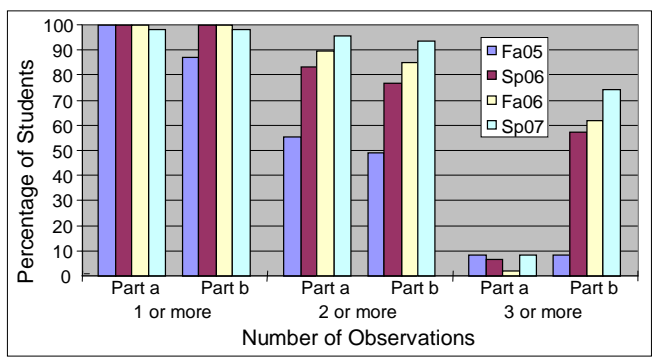}
  \caption{Percentage of students giving correct observations in the second exam question.}
  \label{observations}
\end{figure}

Fig. \ref{observations} shows the percentage of students giving correct observations in each part of the second exam question.  Correct observations for part a include that the number of electrons ejected is proportional to the intensity, that the kinetic energy of the electrons remains constant if the frequency of the light is unchanged, that which electron is ejected is random, and that there is no time delay between the light hitting the surface and electrons being ejected.  Correct observations for part b include that the kinetic energy of the ejected electrons is proportional to the frequency of the light, that there is a cutoff frequency below which no electrons are ejected, and that more electrons are ejected as the frequency increases.  Nearly all the students were able to state at least one correct observation in each part, and aside from Fall 2005, when the wording of the question was less clear, the majority were able to make at least two correct observations in part a and three in part b.  Thus, most students were able to correctly describe the results of the experiments described in this question.

\begin{figure}[bh]
  \includegraphics[width=\columnwidth]{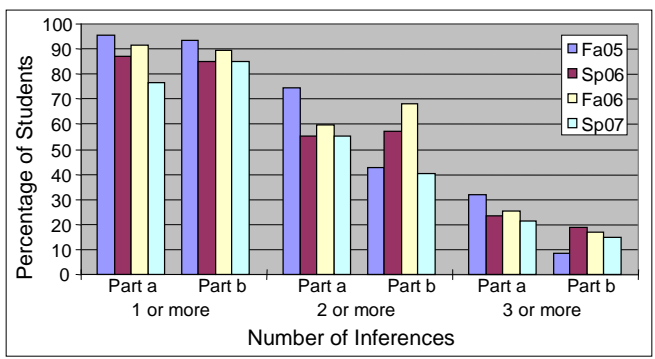}
  \caption{Percentage of students giving correct inferences in the second exam question.}
  \label{inferences}
\end{figure}

Fig. \ref{inferences} shows the percentage of students giving correct inferences in each part of the second exam question.  Correct inferences for part b include that light is made of photons, that the energy of a photon is proportional to the frequency, and that higher energy photons can eject more tightly bound electrons.  Nearly all the students were also able to state at least one correct inference in each part, and on average the majority were able to state at least two correct inferences for each part.  Thus, in addition to correctly describing the observations of the photoelectric effect, students were also able to correctly state the inferences that can be made from these observations.

\begin{figure}[tbp]
  \includegraphics[width=\columnwidth]{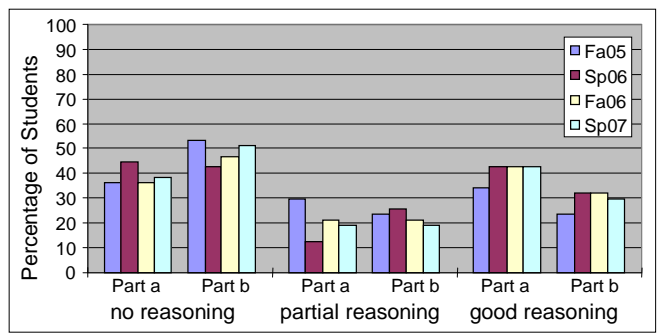}
  \caption{Student reasoning connecting observations and inferences in the second exam question.}
  \label{reasoning}
\end{figure}

Fig. \ref{reasoning} shows student reasoning connecting observations and inferences in each part of the second exam question.  A response was marked as ``good reasoning'' if a student fully explained how at least one correct observation leads to at least one correct inference, ``partial reasoning'' if a student gave some explanation relating an observation to an inference but did not fully explain how one leads to the other, and ``no reasoning'' if a student gave incorrect, unintelligible, or no reasoning, for example simply restating an observation as a reason for an inference.  Only about a third of the students were able to fully explain how the observations led to the inferences.  Further, many students were confused by the difference between observations and inferences, giving inferences in response to questions about observations and vice versa, or responding to part c as if it were simply repeating the question asked in parts a and b.

\begin{figure}[tbp]
  \includegraphics[width=\columnwidth]{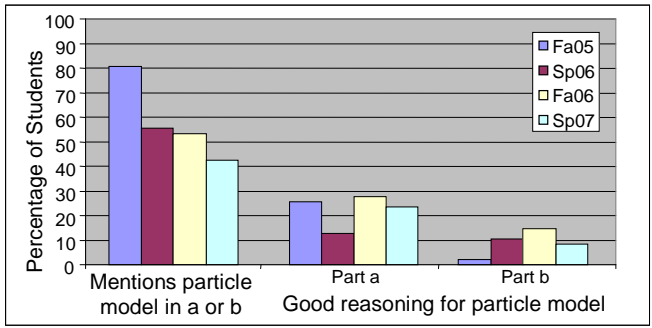}
  \caption{Percentage of students discussing the particle model and giving good reasoning for how the observations lead to the inference of the particle model.}
  \label{particle}
\end{figure}

From an instructor's perspective, the main goal of the second exam question is to determine whether students can correctly reason from observations of the photoelectric effect to the particle model of light.  The first column of Fig. \ref{particle} shows the percentage of students who mentioned the inference that light is made of particles.  Unlike most of the results mentioned in this paper, this percentage varies dramatically from semester to semester, declining from more than $80\%$ in Fall 2005 to less than half in Spring 2007.  One possible explanation for the declining percentage of students mentioning the particle model of light is that the more specific wording of the question focuses the students' attention on details rather than on the big picture.  The second two columns of Fig. \ref{particle} show the percentage of students who gave good reasoning for how the observations led to the inference of the particle model in each part.  These low percentages do not necessarily imply that most students did not realize that light is made of particles or could not explain the reasoning behind this inference if they were asked to do so more explicitly, but that they did not view this idea as relevant to answering this question.

\section{Conclusions}

The analysis of student responses to the first exam question shows that the majority of students using our curriculum have achieved the first learning goal, to be able to correctly predict the results of experiments of the photoelectric effect.  Students did well on all aspects of this question, including the section on the effect of changing voltage, which appears to be the most difficult aspect of the experiment for many students.

The results of the analysis of student responses to the second exam question, designed to test the learning goal of describing how the results of the photoelectric effect experiment lead to the photon model of light, are more ambiguous.  While most students could correctly state both the observations and the inferences involved in the photoelectric effect, they could not necessarily reason effectively about the connection between the two or even distinguish between them.

These results demonstrate that our curriculum provides a significant improvement over traditional instruction, which leads to many students who cannot describe the basic experimental set-up or conclusions of the photoelectric effect, as shown in previous research.  There is still room for improvement in developing students' skills in reasoning from observations to inferences.  We saw the same problem in other parts of the course.  While we emphasized scientific reasoning skills repeatedly throughout our course, the majority of our curriculum was not explicitly structured towards their development.  We believe that the observed student difficulties are symptomatic of a more widespread shortcoming in physics instruction in general in addressing this important skill.  This is consistent with research by Etkina et al.~\cite{Etkina2006a}, which suggests that the development of such scientific reasoning skills requires entire courses to be structured throughout towards this goal.  Because the history of modern physics includes so many relatively simple experiments that demonstrate important new ideas in science, such as the photoelectric effect, this is an area ripe for further research in developing students' scientific reasoning skills.

\section{Acknowledgments}
We thank Noah Finkelstein for allowing us to take data in his course in Fall 2006 and Spring 2007, and Ron LeMaster, the software engineer for the \textit{Photoelectric Effect} simulation.  We thank all the faculty members who took the time to respond to our survey, and the undergraduate learning assistants and graduate teaching assistants who helped with the course.  We also thank the PhET team and the Physics Education Research Group at the University of Colorado.  This work was supported by the NSF, the Hewlett Foundation, and the University of Colorado.  W. Handley was supported by a Colorado STEM Noyce Fellowship.

\bibliography{../bibliographies/PER}
\bibliographystyle{apsrev}
\end{document}